# Toward Lean Industry 5.0: A Human-Centered Model for Integrating Lean and Industry 4.0 in an Automotive Supplier

Peter Hines[a], Florian Magnani[b], Josefa Mula[c]* and Raquel Sanchis[c]

[a]*Business – Management and Organisation, South East Technological University, Waterford, Ireland, 0000-0003-1169-5912;*

[b]*Université Jean Moulin Lyon 3, iaelyon School of Management, UR Magellan, Lyon, France, 0000-0002-5970-2221;*

[c]*Research Centre on Production Management and Engineering (CIGIP), Universitat Politècnica de València, Alcoy, Alicante, Spain, 0000-0002-8447-3387; 0000-0002-5495-3339.*

*Corresponding author: Josefa Mula, fmula@cigip.upv.es

# Toward Lean Industry 5.0: A Human-Centered Model for Integrating Lean and Industry 4.0 in an Automotive Supplier

## Abstract

This paper proposes a human-centered conceptual model integrating lean and Industry 4.0 based on the literature review and validated it through a case study in the context of an advanced automotive first-tier supplier. Addressing a significant gap in existing research on lean Industry 4.0 implementations, the study provides both theoretical insights and practical findings. It emphasizes the importance of a human-centered approach, identifies key enablers and barriers. In the implementation process of the case study, it is considered at group level and model site level through operational, social and technological perspectives in a five-phase multi-method approach. It shows what effective human-centered lean Industry 4.0 implementation look like and how advanced lean tools can be digitized. It highlights 26 positive and 10 negative aspects of the case and their causal relation. With the appropriate internal and external technological know-how and people skills, it shows how successful implementation can benefit the organization and employees based on the conceptual model that serves as a first step toward lean Industry 5.0.

## 1. Introduction

Since the term lean was coined in the late 1980s, it has become the dominant improvement paradigm in manufacturing industry. Lean is a way of thinking that is designed to create a culture in which everyone in the organization continuously improves operations and, hence, efficiency (Found & Rich, 2007). This position has not been rivaled until the advent of Industry 4.0 (I4.0), a term coined at the Hannover Fair in 2011 (Sanders et al., 2016). I4.0 was first illustrated by Kagermann et al. (2011, 2013) as a part of a smart, networked ecosystem leveraging the interconnectivity of products, machines and employees by using systems' growing decision-making capabilities to allow companies to improve their agility and profitability (Rosin et al., 2019). There has been considerable academic debate about how lean and I4.0 (LI4.0) might be brought together (Bittencourt et al., 2021; Pagliosa et al., 2021). LI4.0 was first studied by Dombrowski et al. (2017), who conducted pioneering research into the interdependencies between these two approaches by analyzing a use case that emphasized both process and technical aspects.

LI4.0 can be defined as a hybrid operational approach that integrates lean practices with I4.0 technologies to achieve enhanced productivity, agility and sustainability (Buer et al., 2021). Many recent articles have studied LI4.0 with most involving either literature reviews (Kipper et al., 2020; Pagliosa et al., 2021) or industry surveys (Tortorella et al., 2018). The most comprehensive summary of these calls is provided by Hines et al. (2023) based on the views of the top 300 cited authors in the field. The most important identified research gaps are: lack of clarification of the social perspective (Romero et al., 2020); the identification of enablers and barriers to implementation (Stentoft & Rajkumar, 2020); the clarification of the competencies, training and education required for successful implementation (Ansari et al., 2018); the implementation results (Biondo et al., 2024); the means to combine lean and I4.0 theoretically (Hines & Netland, 2023). In short, an understanding of the mechanics of LI4.0 implementation from the operational, technological and social perspectives is clearly lacking. From an operational perspective, there is a need to recognize what is required at the site level, and for larger organizations from a group view. From a technological perspective, it is necessary to understand how technology leverages the integration and alignment of lean practices with I4.0 technologies. From a social perspective, interpreting how social engagement can be beneficial to successful implementation is another need.

The literature reveals that the conceptual frameworks and models which integrate LI4.0 exist, but do not include the operational, technological and social perspectives at the same time. The research performed by Kaswan et al. (2023) aims to propose a framework to integrate green lean six sigma and I4.0 (LSS 4.0) to improve organizational sustainability, focusing solely on this goal while leaving aside other objectives such as efficiency, continuous improvement, among others. The framework proposed by Citybabu and Yamini (2023) integrates lean six sigma 4.0, however this study does not provide empirical evidence from real implementation or use cases. The same applies to the conceptual framework that integrates (LSS 4.0) (Citybabu & Yamini, 2024), which encompasses operational, sustainability, and human factors or ergonomics aspects, as it is just based on theoretical evidence from the literature review.

Additionally, there is a lack of case studies on LI4.0 even at a site level (Rossini et al., 2019) let alone at a group level (Frecassetti et al., 2024). The limited cases include: the links between lean and I4.0 in the Italian manufacturing industry (Ciano et al., 2021); the significance of I4.0 to improve communications and people recognition to improve lean manufacturing and engagement in the UK automotive industry (McKie et al., 2021); and the linkages between lean and I4.0 in the Italian electronics industry (Dieste et al., 2023).

However, holistic case studies that cover the range of the key gaps identified above are presently lacking. The motivation of this work is to fill these gaps by offering a conceptual model that supports how to implement LI4.0 from a human-centered perspective through studying an in-depth case at both group and model plant levels. This topic is both important and timely given the rapid rise in firms implementing LI4.0 (Pagliosa et al., 2021; Rosin et al., 2019), persistent confusion in industry about its implementation (Hines & Tortorella, 2024; Nayernia et al., 2022) and lack of a strong theoretical foundation for LI4.0 implementation backed with empirical evidence (Ciano et al., 2021). Moreover, understanding LI4.0 is crucial because it serves as the first step toward the Lean Industry 5.0 (LI5.0) paradigm (Fani et al., 2024). Industry 5.0 (I5.0), characterized by its human-centered, sustainable and resilient approach (Breque et al., 2021), is increasingly being paid attention. Our research aims to demonstrate how LI4.0 can serve as a foundational step toward LI5.0 by integrating a human-centered perspective. This transition bridges the LI4.0 approach with Industry 5.0 principles by ensuring a more holistic and human-centered approach to industrial transformation.

This research enables the main following research question to be addressed: (RQ) what should be considered to integrate the triad of lean, I4.0 and human? It can be divided in two sub research questions: (RQ1**)** what mechanisms and/or tools foster this integration? and (RQ2**)** what are the main implementation results of this integration? Hence the main contributions of this paper are to provide not only guidance to managers, but also a theoretical advance in human-centered LI4.0 implementation to the research community.

The remainder of the paper is organized as follows. The literature review explores what research has been conducted specifically in LI4.0 and the implementation theories that have been developed. Then, the proposed human-centered LI4.0 conceptual model is introduced and elaborated. Our multimethod methodology, focused on the case study, is then outlined before the results are presented, which are discussed in the extant literature context. Then we highlight our contributions, conclusions and research limitations.

## 2. Literature review

With regards to the technological perspective, I4.0 suggests a pivotal paradigm shift: companies' value stream entities strategically move toward the digitization of their operational processes by enthusiastically embracing cyber-physical systems to cultivate highly agile and responsive production systems. To date, these companies have concentrated on the implementation of I4.0 technologies to amplify their productivity and monitoring capabilities. Scholars have recognized the increasing need for a synergistic integration of I4.0 principles with lean practices (Salvadorinho & Teixeira, 2020). Lean's intrinsic attributes appear less technology-oriented than I4.0, and focus more on its rigorous process orientation, well-defined tasks and timeframes, work standardization, with an emphasis on visual control and transparency. These attributes effectively facilitate the integration of I4.0's information sharing and automation components (Ciano et al., 2021) by focusing on the operational perspective. It is noteworthy that the literature predominantly scrutinizes lean by focusing on processes and tools, while the I4.0 accentuates a spectrum of technologies or infrastructures (Rossi et al., 2022). However, there is a need to understand how technology leverages the integration and alignment of I4.0 with lean practices, i.e. LI4.0. No research to date has explored whether a specific implementation sequence of LI4.0 optimally enhances synergistic effects (Buer et al., 2021) while strongly emphasizing the social perspective of its integration.

LI4.0 integration effectively manages the interactions among operational (routines and processes), technological (infrastructure and technology) and social (people and culture) perspectives. Yet a singular emphasis placed on either lean's processes and tools (Salvadorinho & Teixeira, 2020) or I4.0's technologies (Tortorella et al., 2023) is deemed insufficient. There has been a tendency to inadvertently overlook the social perspective in this equation, as emphasized by Eslami et al. (2024). It is discernible that the paramount concern continues predominantly to center on optimizing operational efficiency, with the enhancement of employees' working conditions not being accorded equal priority. This highlights the imperative need to clarify LI4.0 from the prism of social perspective, technology acceptance and new forms of autonomy (Rosin et al., 2019; Salvadorinho & Teixeira, 2020). This is also part of the new European Commission Industry 5.0 trend (Breque et al., 2021) that capitalizes on I4.0 technologies with a human-centered focus, and emphasizes collaboration between people and intelligent systems. Unlike its predecessor, which focused on automation and efficiency, Industry 5.0 integrates advanced technologies while ensuring that human creativity, decision making and expertise remain essential (Goujon et al., 2024). By fostering a more inclusive and adaptive work environment, Industry 5.0 seeks to enhance job satisfaction, worker safety and overall social benefits to make industrial transformations more human-centric (Passalacqua et al., 2025).

More recently, extensive research has explored the intricate social perspective of lean, with comprehensive overviews provided by Bouranta et al. (2022). Regarding I4.0 or I5.0, new research shed light to the social perspective from several angles, including collaboration and decision-making processes (Rosin et al., 2019), leadership roles (Tortorella et al., 2018), and new human resources management (HRM) practices, denoted as HRM 4.0 (da Silva et al., 2022). When considering LI4.0, it is noteworthy that the interplay between the social and technological perspectives has not been entirely overlooked. Most current research that addresses the impacts of adopting lean, I4.0 or LI4.0 focus on the measurement of human factors, and also on ergonomics through physical cognitive and psychological effects (Longo et al., 2019; Sakthi Nagaraj & Jeyapaul, 2021). From a social perspective, there is a need to interpret how social engagement can be beneficial to successful implementation.

A stream of research related to the social perspective anchored in the self-determination theory (Ryan & Deci, 2000) elucidates the three core human needs – competence, relatedness and autonomy– as pivotal elements for enhancing the employee experience and, thereby, fostering operational efficiency and employee satisfaction. Consequently, the 'people value stream' notion is introduced, which serves as the human counterpart to the established (product) value stream (Hines, 2022). By drawing from insights into neuroscience and psychology, this people value stream approach is geared to foster individual employees' personal autonomy development, while acquiring more interactional competencies. A systemic perspective is necessary to shed light on the success factors and competencies development related to successful implementation by integrating the rational and irrational aspects of LI4.0 implementation from the social perspective.

Regarding barriers identification, the implementation process and the influential factors of LI4.0 have recently drawn increased attention. From a technological perspective, the barriers to implementation primarily revolve around investment costs and technology readiness, which constitute the most significant hurdles (Hines et al., 2023). From an operational perspective, the difficulty lies in sustaining the routines and new practices that employees and managers perceive as useful (Knol et al., 2018). From a social perspective, top management support, exemplarity and commitment emerge as pivotal success factors for LI4.0 implementation (Samanta et al., 2024; Vinodh et al., 2020). Employees' empowerment is integral to overall implementation success (Yilmaz et al., 2022). Consequently, the establishment of a collaborative learning environment and providing employees with appropriate ways to experiment new technologies are indispensable for achieving implementation success, fostered by team interactions and the constructive tension of creative processes (Saabye et al., 2022).

Recent research digging into the social perspective of LI4.0 implementation advocate for a more comprehensive socio-technical perspective (Marcon et al., 2022). These calls underscore the critical importance of addressing the issues and barriers related to work routines (Åhlström et al., 2021), legitimization (Holweg et al., 2022) and autonomy (Rosin et al., 2019). To further explore these dynamics, Schneider and Sting (2020) advocate exploratory qualitative in-depth case studies that focus on socio-cultural aspects. This entails a keen examination of evolving workers' roles in manufacturing environments to identify the contextual barriers that influence management decisions and contribute to employee acceptance of digitization-induced changes in organizations. Additionally, Olsen and Tomlin (2020) posit that investigating the mechanisms of acceptance or non-acceptance of new technologies at both individual and group levels represents a current gap in the literature and then a natural avenue of inquiry for the operations management field.

To provide a comprehensive overview of the current state of human-centered LI4.0 research, we conducted a search in the EBSCO database using the following query: **TITLE-ABS-KEY ((lean AND industry 4.0) AND (framework OR model OR "literature review" OR theory) AND (human OR people OR employee OR social))**. We reviewed the resulting articles and selected those offering significant contributions from operational, technological, and social perspectives. The final selection (Table 1) highlights the most relevant and integrative models about LI4.0 integration frameworks, and identifies key orientations, barriers, enablers and gaps.

**Table 1.** Overview of the current state of human-centered LI4.0 research.

| | | | Rossi et al. (2019) | Romero et al. (2019) | Rosin et al. (2020) | Salvadorinho et al.(2020) | Åhlström et al. (2021) | Buer et al. (2021) | Ciano et al. (2021) | Hines (2022) | Saabye et al. (2022) | Yilmaz et al. (2022) | Hines et al. (2023) | Samanta et al. (2023) | Vinodh & Shimray (2023) | Eslami et al. (2024) | Gedara & Madusanka (2024) | Salvadorinho et al. (2024) | Sehnem et al. (2025) |
|---|---|---|---|---|---|---|---|---|---|---|---|---|---|---|---|---|---|---|---|
| Orien-tation | | Lean | ✓ | | | ✓ | ✓ | ✓ | ✓ | ✓ | | ✓ | ✓ | ✓ | ✓ | | ✓ | | ✓ |
| | | I4.0 | ✓ | ✓ | ✓ | ✓ | | ✓ | ✓ | | ✓ | ✓ | ✓ | ✓ | ✓ | ✓ | ✓ | ✓ | ✓ |
| | | Human | | ✓ | | | | | | ✓ | ✓ | ✓ | ✓ | | | | ✓ | ✓ | |
| Barriers | Lean | Lack of work routines | | | | | ✓ | | | | | | | | | | | | |
| | | Absence of legitimization | | | | | ✓ | | | | | | | | | | | | |
| | | Lack of autonomy | | | | ✓ | ✓ | | | | | | | | | | | | |
| | I4.0 | Investment costs | | | | ✓ | | | | | | ✓ | ✓ | | | | | | ✓ |
| | | Technology readiness | | | | | | | | | | ✓ | ✓ | | | | | | |
| | Human | Lack of top management support | | | | | | | | | | ✓ | ✓ | ✓ | | | | ✓ | |
| | | Lack of commitment | | | | | | | | | | ✓ | ✓ | ✓ | | | | ✓ | |
| | | Lack of exemplarity | | | | | | | | | | | | | | | | | |
| Enablers | Lean | Continuous improvement | | | ✓ | | | | | | | | | ✓ | | | | | |
| | | Work standardization | | | | | ✓ | | ✓ | | | | | | | | | | |
| | | Control and transparency | | | | ✓ | | | ✓ | ✓ | | | | | | | | ✓ | |
| | | Creative processes | | | | | | | | | ✓ | | | | | | | | |
| | I4.0 | Flexibility | | | ✓ | ✓ | | | | ✓ | | | | ✓ | ✓ | | | | |
| | | Scalability | ✓ | | | | | | ✓ | | | | | | | | ✓ | | |
| | | User-friendliness | | | | | | | | | | | ✓ | | | | | | |
| | Human | Employee's empowerment and autonomy | ✓ | | | | | | ✓ | ✓ | | ✓ | | | | | | | |
| | | Collaborative and learning environment | ✓ | | | | | | | | ✓ | ✓ | ✓ | ✓ | | | | | ✓ |
| | | Team interactions | | | | | | | ✓ | ✓ | ✓ | | | | | | | | |
| Gaps | Lean + I4.0 | Implementation sequence | | | ✓ | | | ✓ | | | | | ✓ | | | | | | |
| | | Alignment between lean tools and digital solutions | | | | | | | | | | | | | | | | | |
| | I4.0 | Human-technology interaction | | ✓ | | | | | | | | | ✓ | ✓ | | | ✓ | | |

| | | | Rossi et al. (2019) | Romero et al. (2019) | Rosin et al. (2020) | Salvadorinho et al.(2020) | Åhlström et al. (2021) | Buer et al. (2021) | Ciano et al. (2021) | Hines (2022) | Saabye et al. (2022) | Yilmaz et al. (2022) | Hines et al. (2023) | Samanta et al. (2023) | Vinodh & Shimray (2023) | Eslami et al. (2024) | Gedara & Madusanka (2024) | Salvadorinho et al. (2024) | Sehnem et al. (2025) |
|---|---|---|---|---|---|---|---|---|---|---|---|---|---|---|---|---|---|---|---|
| | | Social engagement in digital transformation processes | ✓ | ✓ | | | | | | | | | | | ✓ | | ✓ | | |
| | Human-centered LI4.0 | Lack of human-centered LI4.0 approaches | | ✓ | | | | | | | | | ✓ | | | | ✓ | | ✓ |
| | | Consideration for social impact in LI4.0 methodologies | | ✓ | | | | | | | | | | | | | ✓ | ✓ | |
| | | Lack of holistic integration | | ✓ | | | | | | | | | ✓ | | | | ✓ | | |

## 3. Human-centered lean Industry 4.0 conceptual model

Some conceptual models that combine lean with I4.0 have been developed to facilitate their integration, and each one presents distinct challenges (Table 2). For instance, Sony (2018) introduced an integrative model that incorporated lean principles across the three I4.0 integration dimensions: vertical, horizontal and end-to-end engineering. Although this model provides a strong theoretical foundation, it remains largely unvalidated in empirical settings and overlooks the critical role of human factors. Tortorella and Fettermann (2018) highlighted the maturity level of a company's lean system as a key determinant in the successful integration of I4.0 by emphasizing the foundational role of existing lean practices in I4.0 transformation. This was confirmed by Rossini et al., (2019), who underpinned the idea of a wide applicability of both approaches, and indicated that higher I4.0 adoption levels could be easier to achieve when lean production practices are extensively implemented in the company. Bittencourt et al. (2021) reinforced this perspective with a systematic literature review, and confirmed that lean not only facilitates I4.0 adoption, but also creates the need for new technologies implementation. Their findings indicated that research has primarily focused on how I4.0 technologies can support and enhance existing lean practices.

**Table 2.** Comparison of LI4.0 conceptual models.

| Conceptual model | Key elements | Challenges | References |
|---|---|---|---|
| I4.0 architecture | Connectivity among structural elements, information exchange and decision making | Complexity in integration, difficulty maintaining robustness against failures | (Citybabu & Yamini, 2023; Nounou et al., 2022; Sony, 2020) |
| Lean 4.0 impact analysis | Operational performance, alignment with objectives, sequence of adoption | Need for careful alignment of lean practices and I4.0 technologies | (Citybabu & Yamini, 2024; Dillinger et al., 2022; Rosin et al., 2019) |
| Integration framework | Systematic adoption, improved material and information flows, flexibility, mutual facilitation | Complexity in implementation and ensuring seamless integration, overcoming trade-offs among competitive objectives | (Ding et al., 2023; Rossini et al., 2022; Sony, 2018; Vigneshvaran & Vinodh, 2023) |
| **Lean 5.0 paradigm** | Human-centric lean practices, mostly developing engagement and empowerment | Imperative to integrate a human-centric perspective, deeply respectful and appreciative of the human contribution | (Fani et al., 2024) |

Overall, these conceptual models underscore the synergistic relation between lean practices and I4.0 technologies by contributing to greater efficiency. However, significant challenges remain, particularly concerning cultural transformation, organizational alignment, and complexity adoption. These challenges are predominantly tied to the social perspective of implementation. With the exception of the Lean 5.0 paradigm, most existing models relegate the human dimension to a secondary role. However, Lean 5.0 primarily focuses on the interaction of three key technologies, such as cloud computing, the Internet of Things (IoT) and big data; and human-centered lean practices, such as engagement and empowerment (Fani et al., 2024). We build upon the definition provided by Fani et al. (2024), who describe Lean 5.0 as “an emerging paradigm that integrates lean methodologies with advanced human-centric and digital principles. It combines the efficiency and waste-reduction focus of lean practices with the technology-driven enhancements of I4.0 while further enriching this integration by prioritizing human aspects”. Our model serves as a foundational step toward the Lean 5.0 paradigm, or LI5.0, by further enhancing its human-centered features to strengthen the synergy between lean practices and I4.0 technologies. Here we propose the human-centered LI4.0 conceptual model (Figure 1) based on three perspectives: (i) operational, following a lean approach (Rossi et al., 2022), which seeks well-defined processes (Ciano et al., 2021), elimination of waste, and the continuous improvement (Found & Rich, 2007; Kaswan et al., 2023); (ii) technological, following I4.0 aspects (Rossi et al., 2022), based mainly on digitization and automation (Ciano et al., 2021); and (iii) social, human-centered, which focus on addressing human needs, such as reducing physical and mental effort, supporting employees health (Hines, 2022) and encouraging proactive participation.

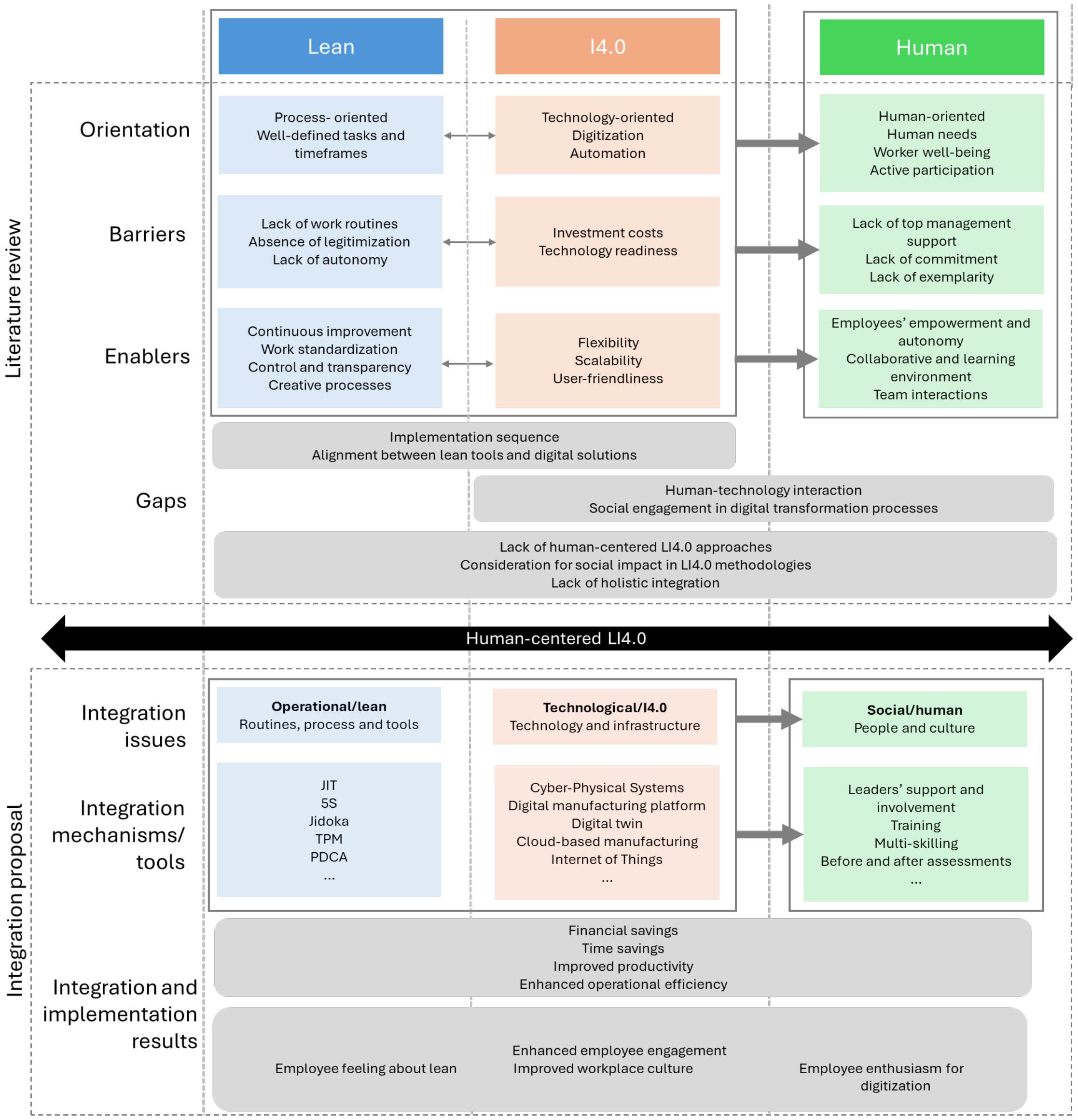

**Figure 1 Caption.** Human-centered LI4.0 conceptual model.
**Figure 1 Alt Text**. Conceptual model involving the integration of lean principles, I4.0 technologies and a focus on human factors with the objective of improving operational efficiency, agility, and profitability, while maintaining the well-being and active participation of employees.

The integration of human-centered LI4.0 presents complexities from the three aforementioned perspectives. The first set of lean barriers includes the lack of structured processes (Åhlström et al., 2021), absence of legitimization (Holweg et al., 2022), and limited autonomy (Rosin et al., 2019), which hinder operational efficiency and workers' engagement. From the technological perspective, the challenges focus on high investment costs and technology maturity and readiness (Hines et al., 2023), factors that can slow down the adoption of necessary I4.0 solutions. Finally, the absence of senior-level support, workers' resistance, and insufficient demonstration of best practices (Samanta et al., 2024; Vinodh et al., 2020) further intensify these barriers, preventing effective leadership and alignment with the integration goals. To overcome these barriers, the necessary lean tools to be integrated for a specific application should be identified; however, continuous improvement, governance and transparency, work standardization (Ciano et al., 2021) and innovative methods (Saabye et al., 2022) are key integration points. Similarly, technology should be tailored to each case, but flexibility, scalability, and user-friendliness (Ojha, 2023) can be enablers for integration. Finally, from a human perspective, aspects such as autonomy and decision-making power (Yilmaz et al., 2022), a collaborative and developmental environment, and team collaboration (Hines, 2022; Saabye et al., 2022) are crucial for integration, as they foster greater adaptability and commitment to human-centered continuous improvement processes and technology integration.

Based on the background elements of the conceptual model presented earlier, the main gaps are identified. To our best knowledge, there is a lack of holistic integration implementations. Such implementations should not only be focused on traditional engineering and computer science knowledge but also give considerable attention to the human resources aspect. Specifically, it is important to address how the affected individuals feel and think before, during, and after the implementation. Moreover, integrating LI4.0 with a human-centered approach is not easy and can be a very confusing journey, without knowing exactly what to implement first and what to implement next, i.e., the implementation sequence (Buer et al., 2021), as well as the alignment between lean techniques and I4.0 technological tools. Continuing with the alignment, gaps have also been identified concerning human-technology interaction, while social management with lean techniques seems better. Consequently, the absence of human-centric I4.0 approaches (Eslami et al., 2024; Romero et al., 2020) in real-world applications highlights the need for a conceptual model validated through a real case study.

Here the human-centered LI4.0 conceptual model aims to address the multifaceted challenges of integrating lean practices with I4.0 technologies in a way that considers operational routines, technological infrastructure, and social dynamics. Considering, as an example, this sequence (i) implementation of lean; (ii) implementation of I4.0 technology; (iii) integration of lean techniques and I4.0 and (iv) integration and focus on the human factor, the conceptual model encompasses various integration mechanisms and tools from the lean perspective such as just-in-time (JIT), 5S, Jidoka, total productive maintenance (TPM), and plan-do-check-act (PDCA), among others. It also includes advanced technologies like cyber-physical systems, digital manufacturing platforms, digital twins, cloud-based manufacturing, and the IoT, to mention a few, for I4.0 implementation. However, the sequence will depend on the specific situation of each real case, as some may already have certain lean techniques implemented, or conversely, the case may be technologically advanced but inexperienced in lean processes (Hines &

Tortorella, 2024). Therefore, the conceptual model does not specify the sequence but rather the mechanisms to be used. Although it is worth noting, based on the statement made by Pagliosa et al. (2021), that I4.0 technologies can enhance the lean approach, which suggests that lean should be implemented first and then improved through I4.0.

As the ultimate goal of the model is to be human-centered, the LI4.0 conceptual model must consider mechanisms like leaders' support and involvement, training, multi-skilling, and assessments before and after implementation. By focusing on these elements, the human-centered LI4.0 conceptual model aims to achieve significant integration and implementation results, including financial and time savings, improved productivity, enhanced operational efficiency, and a positive impact on workplace culture and employee enthusiasm.

## 4. Case study development

According to Stake's (1995) typology, research is founded on an intrinsic case study. The authors did not take part in the implementation of the case and were, therefore, independent. Hence work did not suffer from any biases associated with action research. It is particularly well-suited for exploring the integration of lean practices and I4.0 technologies because it allows for an in-depth, context-rich examination of a specific case in which this convergence is taking place. This methodology is justified for three key reasons: (i) the complexity of LI4.0 integration; (ii) the context-specific operational and technical challenges; (iii) the human-centered implications of the implementation.

The case company had adopted lean over 20 years ago and was a pilot for the implementation of I4.0 in the group. This implementation of I4.0 technologies was across the board and involved seven platforms, which make it advanced in LI4.0 and, hence, an ideal case.

The company itself is the seventh largest global automotive supplier with around 300 plants and 150000 employees worldwide. The main part of the business centers on car interiors, electronics and clean mobility, with another part focusing mainly on lighting products. There is a marked focus on customers with the company maxim of 'total customer satisfaction: everyone, everywhere, every time'. The corporate facility is in France, whereas the pilot site that we focus on this research is in Spain.

This section presents the validation methodology for a human-centered LI4.0 model plant that integrates operational, social and technological perspectives.

### *4.1. Validation methodology*

Much of LI4.0 research has been undertaken by literature reviews or quantitative industry surveys (Núñez-Merino et al., 2020). The widespread use of this latter method in LI4.0 research potentially to lead to repetition and inward reflection is of particular concern because quantitative surveys can suffer from bias given the way that samples are obtained (Guide Jr. & Ketokivi, 2015).

In this research, the authors choose an evolutionary qualitative approach by employing a single exemplar case study supported by information from the organization's group level (Figure 2). The researchers wish to explore an effective way of integrating lean and I4.0 from operational, technological and human perspectives to gain an holistic picture of what is happening, what the acceptance mechanisms were and why.

This largely qualitative research is supplemented by 'before and after' quantitative questioning to establish impacts, where two tailed t-tests are applied to test the significance of the results. This mixed method approach allows for triangulation (Maxwell, 2013), which means that we can gain secure understanding (Fielding & Fielding, 1986). It also allows us to secure information about different aspects of the case by fostering complementarity and expansion (Greene, 2007). This is particularly important to acquire tacit understanding of areas participants might have been reluctant to reveal during interviews.

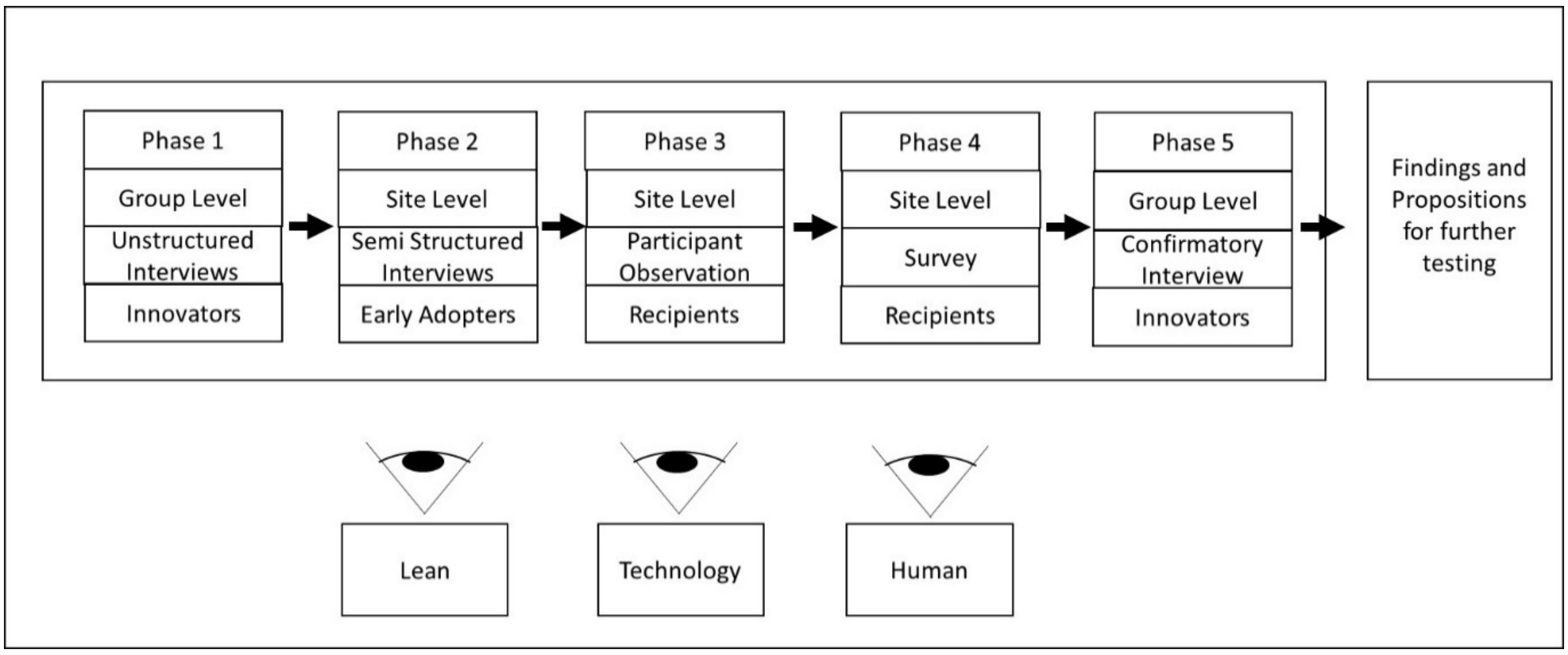

**Figure 2 Caption.** Validation phases.
**Figure 2 Alt Text.** Data capture and verification in research passes through five phases based on: unstructured interviews, semi structured interviews, participant observation, survey and confirmatory interview.

There is progression from a grounded theory exploratory approach to an explanatory one while confirming the findings. The initial grounded theory exploratory approach is very flexible and open, and allows us to explore what occurs at the group and site levels without any preconceptions about causality between operational, social and technological perspectives. To help understand what we find, we follow Saabye & Powell's (2024) case research in LI4.0 when using thematic coding (Braun & Clarke, 2006). This allows us to narrow our focus before conducting semi-structured interviews and later structured explanatory research. This is necessary as the design of this latter research may not have been determined without prior exploratory research (Light et al., 2009). It also enables us to become familiar with the organization's specific terminology and the meaning of actions and events from the people involved and the perspectives that inform these actions (Fielding & Fielding, 1986).

The initial exploratory research is followed up by more explanatory activity to reinforce the replicability of the case organization in its rollout or for other organizations.

It permits us to make our understanding more profound and to provide the opportunity to understand local causality at a constructs level and with the themes within it (Miles & Huberman, 1984).

We follow Cimini et al. (2023) in employing a causal loop diagram (CLD) to shape themes into variables and to plot the relation between variables. CLDs help us to represent dependencies and feedback processes in quickly capturing hypotheses about the causes of dynamics in systems (Sterman, 2002). These types of system thinking models have been adopted by others to assess improvements in company (Martinez-Olvera, 2010), productivity (Kamble & Wankhade, 2021) and workforce performance (Alefari et al., 2020).

Finally, our research approach becomes confirmatory so that we can test our propositions with the case company's group level allowing for an extra stage of validation and rigor (Butler, 2014). Each research methodology phase is detailed below.

The first phase involves a series of five unstructured interviews over a 3-month period with three corporate team members at the case company based on France. The participants are the global head of LI4.0 and two people, who directly report to the head, are innovators in the global program and are all highly experienced in LI4.0. The purpose of these interviews is to understand the group LI4.0 program. It also helps the researchers to understand industry- and firm-specific language and terminology. The use of a series of interviews is designed to avoid single-interview non-disclosure and non-recall, to improve accuracy and to provide opportunities for verification and reflective questions (Creswell, 2012). The unstructured approach is chosen to gain a clear understanding and to allow for unforeseen answers (Adeoye-Olatunde & Olenik, 2021).

The second phase involves employing a series of semi-structured interviews with five senior staff members at the Spanish pilot site (Appendix A). These individuals can be regarded as ‘early adopters’ in the global roll out of the program and the leaders of it at the site level. They include the plant director, a supervisor and the three sites-based I4.0 implementation agents. They are all highly experienced in lean practices and, although starting out with limited knowledge of I4.0 technologies, they have all now become experts. The purpose of these interviews is to understand what happens at the site level, the site-specific terminology, the role of these individuals and to gain insights into the program’s effectiveness. Semi-structured interviews are selected because they permit researchers to be focused, while still conferring the autonomy to explore pertinent ideas that may come up during interviews, hence further enhancing understanding of the case (Adeoye-Olatunde & Olenik, 2021).

The third phase involves shop floor level participant observation to corroborate and validate what had been found out in the first two phases, and to identify missing information or unexpected findings. The researchers are able to use their eyes and ears as tools to collect information and make sense of it (Maxwell, 2013). During this tour, six front-line staff members (supervisors, gap leaders, and operator levels) are questioned to provide the respondent validation of the information gleaned in Phase 2.

The fourth research phase is more explanatory in nature and consists of an anonymous qualitative electronic survey of what might be regarded as the program’s ‘recipients’ (Appendix B). These include one manager, five supervisors, eight gap leaders (team leaders), three indirect staff members and five operators. These were all experienced lean practitioners with initially limited knowledge of I4.0. The purpose is to gain a wider personal perspective on the program from those that are at the receiving end of the work.

The last work phase is to test the consistency of the findings developed with the group level senior team. This is undertaken with two group staff members. These checks are made to rule out the possibility of misinterpreting the participants' meaning, their perspectives as to what is going on, and to help identify any biases and misunderstandings of what has been observed (Maxwell, 2013).

These five phases intend to minimize any biases in our research design and to, hence, ensure overall rigor and trustworthiness due to the reflexive subjective nature of qualitative research (Galdas, 2017). We attempt to implement verification strategies integrally into our research design with several self-correcting mechanisms while researching. This involves a combination of a multimethod approach, as well as several validation and corroboration mechanisms, during research, as outlined in Table 3.

**Table 3.** Control of potential research biases.

| Bias | Description | Potential problem | Countermeasure in research design before research | Self-correcting countermeasure within research |
|---|---|---|---|---|
| Hawthorne effect or participant reactivity | People's tendency to behave differently when they become aware of being observed | Distortion of conclusions, especially about the causal relations among variables | Use of semi-structured interviews and qualitative surveys to mitigate the potential Hawthorne effect on participant observation | Multiple participant observations, verification conversations with participants and key user guides |
| Observer bias | Researchers' expectations, opinions or prejudices influence what they perceive or record in the study | This may lead researchers to note some observations as relevant, but ignoring other equally important observations | Use of semi- structured interviews and a qualitative survey to mitigate the potential observer bias in participant observation | Use of many researchers during unstructured interviews and validation conversations by 2nd and 3rd researchers around semi-structured interviews and participant observation |
| Recall bias | Systematic difference in the participant groups' ability to accurately recall information | It threatens the internal validity and credibility of studies that use self-reported data | it focuses on the recent strong digitization of the factory (< 2 years), use of many interviews of the organization's levels and participant observation to validate the participants' recall information | Corroboration of findings that seem unusual with further discussions during future interviews and participants during participant observation |
| Social desirability bias | Respondents answer questions, which they believe will make them look good to others, but conceal their true | This can affect the validity of behavioral research because people can act in a socially appropriate way | Use of anonymous surveys to validate interviews and surveys, and allowing respondents to complete them at a time and place | Use of participant observation to support interviews, especially where claims appear to be possibly exaggerated |
| | opinions or experiences | and may seek to tell overly positive results | of their choice where they can be undisturbed | |

### *4.2 Operational and social integration*

The group's approach to LI4.0 is first viewed through its lean perspective, specifically the company excellence system or company-specific production system (XPS) (Netland, 2013) which is its operational lever for its people to impact customer satisfaction and ensure a sustainable profitable company. It was designed to create a shared work culture with decentralized organizational principles, effective people development and performance management (Figure 3).

At its base is the building block of people development and stable conditions and, sitting on this, are the three pillars of just-in-time (JIT) conditions, built on quality and efficiency with a cap stone of drive improvement. The people development and stable conditions block is the starting point for the improvement process, and it focuses on: engaging people; developing their competencies in autonomous teams; establishing a safe working environment; maintaining harmony with nature; taking care of the planet. This sustainability also relies on standardizing each step of production and leveling activities.

The JIT conditions block is in place because the company recognizes that it needs to reduce waste during operations, especially by reducing overproduction and lead times. The built-in quality block is in place because customer satisfaction is built on their perception of their experience, particularly product quality, and such quality from the beginning. The efficiency block is designed to improve productivity by reducing variation and waste, and decreasing non value-adding activity. Last, the drive efficiency block focuses on the leadership of operations in driving plant performance with continuous improvement kaizens and breakthrough projects, while ensuring sustainability through routines and, hence, guaranteeing that everyone is moving in the right direction.

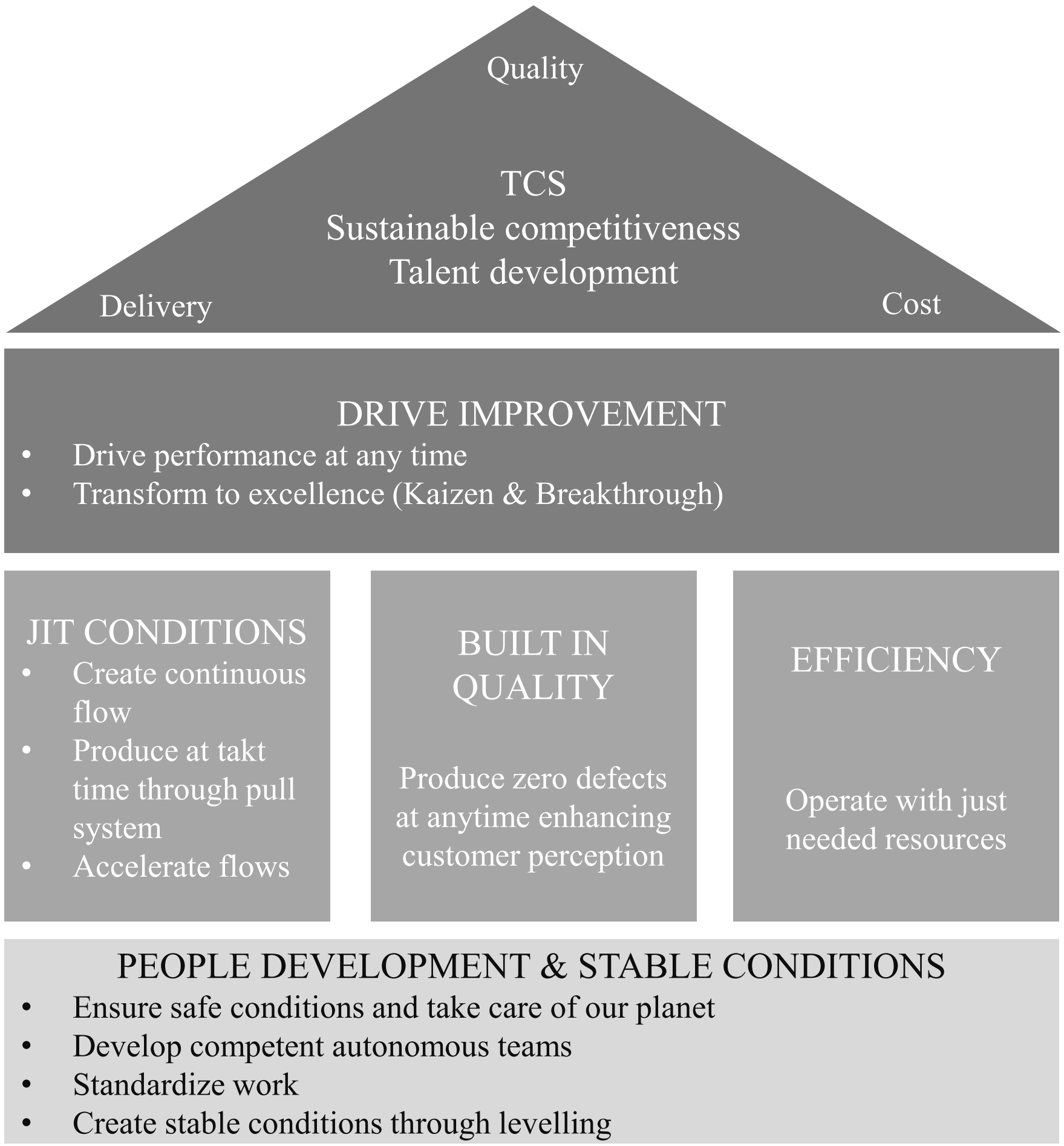

**Figure 3 Caption.** XPS, house of excellence (based on an internal document of the company).
**Figure 3 Alt Text.** XPS of the company studied is based on a house of excellence, designed to deliver customer satisfaction with five building blocks.

Sitting inside these five domains is a set of 11 principles that form the framework for the deployment of XPS in 61 elements or process areas, such as recruitment, small batch single minute exchange of die (SMED), supplier management, automation excellence and problem solving. All these elements are given an owner. There is a wide-ranging education approach to disseminate XPS, including a handbook, an audit, a training matrix, a site best practice database, a learning management system, a digital learning platform or a massive online open course (MOOC), regional workshops for leaders and a daily push flash message to help to embed learning with a suggestion for further training if people do not correctly answer the daily question. The handbook, which is available as both a hard copy and an app, describes what is expected in each fundamental, "does and don'ts", and links with further training materials. The learning management system pushes digital individual learning to employees based on their position in the training matrix. The MOOC sets a basic level of training and certification. There are two further levels, the next being at an advanced MOOC level which is in groups and involves experiential learning. At the highest level are the workshops that involve peer validation. For each organization level, there is a defined profile of which level is needed to be reached in each fundamental.

### *4.3 Technological and social integration*

The company is currently using a digital manufacturing platform, which includes three main set of I4.0 technologies: smart automation driven more locally by a business group; a digital production system driven from a group; a digital platform and artificial intelligence (AI) driven partly by a group and partly by a business group. The introduction of I4.0 technologies started in 2017 mainly focusing on smart automation (860 smart robots, 1100 AGVs) and on Digital Production System. The smart automation focuses on solutions for automated receiving, kitting and picking, smart robots and automated guided vehicles. The digital platform and AI category are a cluster of data-driven solutions for quality inspection, process stability and enhanced manufacturing process inspection. The digital production system includes solutions for strategy deployment and action planning (eKPI and eTask assignment), dynamic production planning (DPP), production monitoring or manufacturing execution system (MES) and planning (eKanban), logistics tracking (eLogistic), as well as the capability to add further solutions, such as augmented workstations and demand variability forecasting.

The company accelerated the I4.0 adoption by adding more robots (250) and AGVs (380), by developing their Digital Production System (30 new model plants, labor management software), and by introducing Data Platform and AI (14 uses cases). This enabled between 40 to 60 M€ savings per year until 2024. The plan for next year, based on feedback from employees, middle managers and directors, is to increase the perimeter of the Digital Production System (150 model plants, real time plant control tower) and to leverage the data generated by the digital production system.

Cloud-based manufacturing integrated planning (CMfg), which has been applied to 15 of the 20 lines, is based on an enterprise requirement planning (ERP) module and uses a spreadsheet interface. It works well, but is quite slow, and this limits the ability to do scenario planning. DPP, also applied to 15 lines, is a solution that is typically run once a day and provides the schedule to fill the leveling board and to, hence, export data to eKanban, a real-time digital sequenced pull system with electronic kanbans, applied in 47 of the 53 work areas. It manages the whole information flow by replacing the need to manipulate kanban cards. Picking lists are provided on a tablet and orders are picked with bar code scanners. Production kanban information is provided on machine side launcher screens.

MES provides the real-time tracking of all 65 machines on the eParts digital board next to them for deviations from standard of the overall equipment efficiency (OEE), quality through statistical process control (SPC) and labor efficiency. It provides a gradually escalating alert (including a line stop signal) so that these can be addressed in a similar way to physical short interval control boards. This information is also reviewed daily, weekly and monthly *via* tablet-based business intelligence (BI) reporting with Pareto ranking of issues. There are also line status boards that display in real time the OEE from each line, whether they are on target and the time to the next changeover.

Plants are provided with 3-month on-site support for the implementation of these technologies with scaling up on their own over the next 3 months.

Noticeably, they were able to digitize their lean practices. A 12-function app suite for management with the local gap teams (eLean) was developed to provide an intuitive shop floor toolbox to support problem solving with the twin aim of improving performance and people development. It particularly addresses gap leaders (team leader) and supervisors (of several gap leaders), and its uptake involves training over a 2-week period. This includes the following solutions: routines for the supervisor and gap leaders; real-time status updates with alerts; real-time action planning and input of improvement actions; local routine coaching; a visual standard work routine, including videos of each role played by gap leaders for employee training; audit of standard work by the gap leader; a multiskilling matrix in which employees can improve their attainment with a rotation matrix to reduce repetitive strain injuries. It also included alerts for retraining; guided improvement of seven local waste types; self-assessment of XPS, including what needs to be developed; a local audit system; customized local total productive maintenance activity plans; and guided 5S activity, audit and tracking.

### *4.4 Operational, technological and social integration*

The LI4.0 model plant is a set of solutions inherent to the lean digital production system including a strong social component (Figure 4). The first is production planning, scheduling and execution, within which their sales and operations and master production schedule (MPS) are managed by CMfg; fine scheduling and leveling are managed by DPP, which schedules finished goods production with upstream picking order and production orders digitally pulled by eKanban. The second is production efficiency and quality building. Here production control and production efficiency are managed by MES that, like eKanban, was internally developed. MES counts the number of produced parts and the scrap rate so that OEE scores can be calculated in real time. The third, eLean, is a suite of lean tools, which was developed together with a local startup business.

The model plant program is designed to be piloted at three plants in Spain, Portugal and the Czech Republic starting in 2020 with a view to a roll out to 30 further 'mother' plants and then in 30 plants per year from 2022 onward. The model plant discussed here is the Spanish one, which employs 435 people, and where the main implementation time period goes from late 2020 to the end of 2021. The enablers for this implementation are the real-time connectivity of machines, the cleaning of the master data, and the appointment of a full-time plant level digital leader and four solution champions, who all spend 50% of their time.

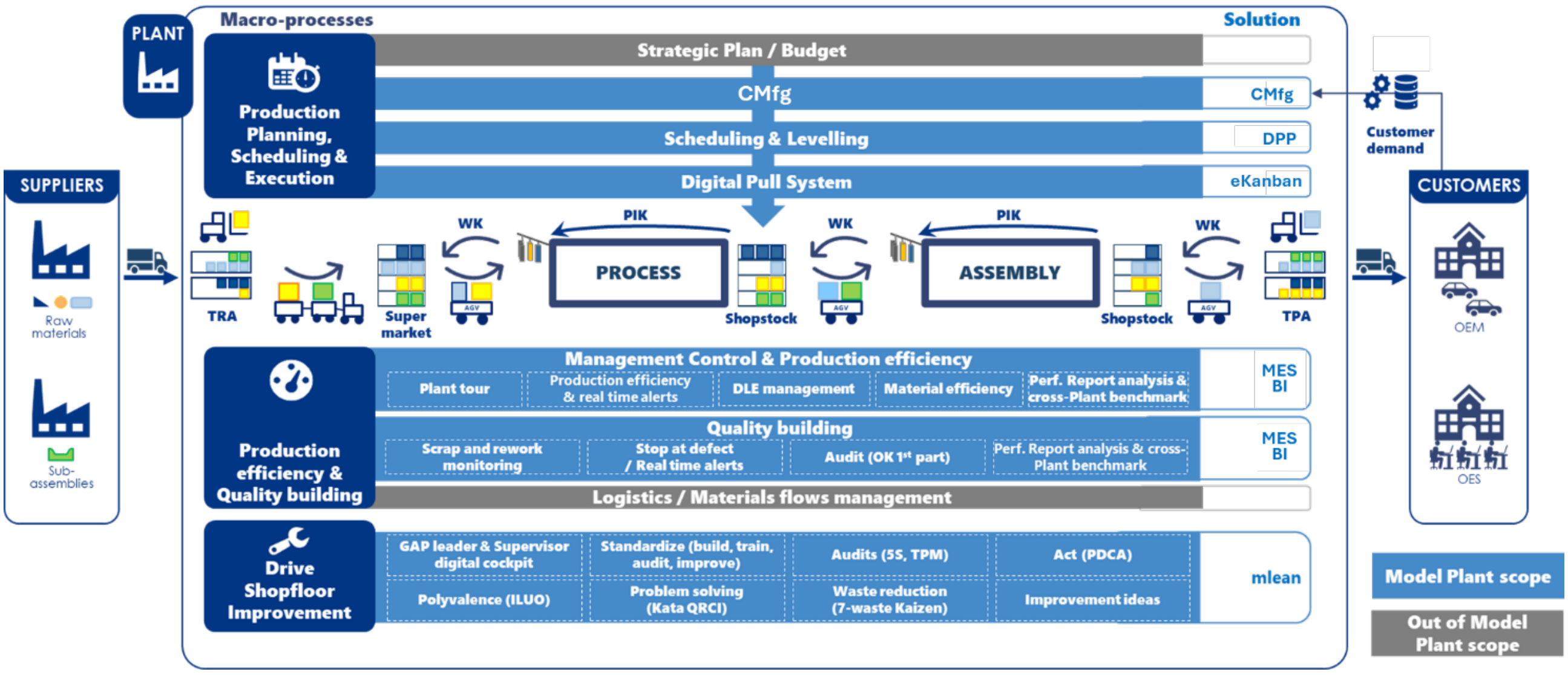

**Figure 4 Caption.** Model plant overview (based on an internal document of the company).
**Figure 4 Alt Text.** Model plan consists of three macroprocesses: production planning, scheduling and execution.

As in other lean applications, perhaps the most important role is that of the direct team leader or gap, who leads a group of 4-6 people. This role has been digitized in eLean, including the management of KPIs and routines (TOP5), checking off first part production (OK first part), and the checking key points and meetings in their daily routine (Plant Tour), training (Training) and multiskilling (Polyvalence), plus six other apps.

Thus, in the model plant, the XPS house of excellence is populated with a suite of solutions that are either from proprietary vendors that are developed in-house or with a start-up supplier. Each plant can self-assess where it is on this excellence journey with the eXPS app.

### *4.5 Results Analysis*

In order to identify the effectiveness of lean and the feeling of people about work from a social perspective, several interviews were carried out with the site's leadership level, and those involved in implementing and managing new technologies, who we call 'early adopters', with five respondents (*n*) (Table 4). This is supplemented by observations and a survey of 'recipients' of the technology based on the shop floor. We divide this group into two parts. The first group of 17 we call 'leader recipients', who are a cluster of people involved in managing others, including office staff, supervisors and gap leaders. The second group includes five shop floor 'operator recipients'.

Most of this research is qualitative in nature, with the exception of several quantitative questions employing a seven-point Likert scale whose answers are summarized in Table 4. The first of the questions concerns respondents' views of the effectiveness of lean before and after digitization work, i.e. the integration of operational and technology perspectives from the social viewpoint. In all three groups (early adopters, leader recipients and operator recipients) there is a significant (0.05 level) increase in their perception of lean effectiveness with the largest increase recorded by early adopters (4.30 to 5.90) and operator recipients (4.40 to 5.80), with a smaller increase for leader recipients (4.47 to 5.59). To gain a more emotional response for the new digital technology, the 27 participants were asked about how they felt about their work before and after digitization (Table 4). Once again, in all three groups there was a significant (0.05 level) increase in how people felt after digitization. The largest increased recorded was by operator recipients (4.40 to 5.80), followed by early adopters (4.80 to 5.70), with the least increase in leader recipients (5.00 to 5.35).

In all, 22 people indicate an improvement, three pointing out something worsening and one informing the same. All the last four are from the leader recipient group (2 office workers, 1 supervisor and 1 gap leader).

**Table 4.** Quantitative survey responses.

| | | Early adopters | Leader recipients | Operator recipients |
|---|---|---|---|---|
| | *n* | 5 | 17 | 5 |
| Effectiveness of lean | Before digitization | 4.30 | 4.47 | 4.40 |
| | Standard deviation | 1.04 | 1.12 | 0.55 |
| | After digitization | 5.90 | 5.59 | 5.80 |
| | Standard deviation | 0.74 | 0.71 | 0.45 |
| | Significance level for change | 0.05 | 0.05 | 0.05 |
| Feeling about work | Before digitization | 4.80 | 5.00 | 4.40 |
| | Standard deviation | 0.76 | 1.12 | 1.10 |
| | After digitization | 5.70 | 5.35 | 5.80 |
| | Standard deviation | 0.84 | 0.93 | 0.45 |
| | Significance level for change | 0.05 | 0.05 | 0.05 |
| | Enthusiasm for lean digitization | 6.80 | 5.47 | 6.00 |
| | Standard deviation | 0.40 | 1.20 | 0.00 |

All the respondents were asked the reason for change in views, with 46 positive responses and eight negative responses (Table 5). There are four main positive areas. In two of them, information availability and reduction in workload, the primary impact can be seen to be in terms of increasing the autonomy of the people at the site. In the other two, improved operations and easier to use, the primary impact lies in improving the involved individuals' competence. The bigger number of responses per person from the early adopter group can be explained by the fact that they were interviewed, whereas the other groups were surveyed. Apart from one comment made by an operator, all the negative points are raised by leader recipient group members. There are several viewpoints about lack of training and increased workload, particularly if new technology does not work as well as planned.

**Table 5.** Major themes in the change of perception of lean effectiveness with the digitization of the model plant.

| | Aspect | Times raised by early adopters | Times raised by leader (recipients) | Times by operator (recipients) | Total times raised | Primary motivational impact |
|---|---|---|---|---|---|---|
| Positive | Information availability | 3 | 8 | 2 | 13 | Autonomy |
| | Improved operations | 1 | 6 | 4 | 11 | Competence |
| | Easier to use | 3 | 4 | 1 | 8 | Competence |
| | Reduction in workload | 5 | 3 | 0 | 8 | Autonomy |
| | Faster | 0 | 2 | 0 | 2 | Competence |
| | Collaborative | 0 | 1 | 0 | 1 | Relatedness |
| | Better problem solving | 1 | 0 | 0 | 1 | Competence |
| | Communication | 1 | 0 | 0 | 1 | Relatedness |
| | Predictive | 0 | 1 | 0 | 1 | Competence |
| | Total | 14 | 25 | 7 | 46 | |
| Negative | Increase in workload | 0 | 1 | 1 | 2 | Autonomy |
| | Lack of training | 0 | 2 | 0 | 2 | Competence |
| | Difficult to change habits | 0 | 1 | 0 | 1 | Autonomy |
| | No positive results so far | 0 | 1 | 0 | 1 | Competence |
| | Cost a lot | 0 | 1 | 0 | 1 | Autonomy |
| | Information unavailability | 0 | 1 | 0 | 1 | Autonomy |
| | Total | 0 | 7 | 1 | 8 | |

Participants' responses are shown in Figure 5. All the comments made by early adopters were positive or neutral, with the respondents evenly split between feeling- or thinking-based responses. Two responses, 'higher motivation' and 'proud', suggest that growth needs are being met, while one response, 'less frustrated', denotes that deficiency needs are being met. Two responses, 'good' and 'easier, are aligned with effectiveness, and two address thinking-based technical issues: 'less data processing' and 'logical. In the leader recipient group, similarly there is a set of positive comments that ranges from those addressing higher level emotional needs to thinking-based technical needs. In addition, there is a series of negative comments that goes from emotional responses: 'insecure' (negatively impact on growth needs) and 'suffer' (negatively impacting deficiency needs) to negative responses about effectiveness 'hard' and technical needs 'more manual work'. There are only three comments from the operator recipients that are all emotional in nature: 'more access to knowledge' (growth need) and 'like' and 'more comfortable' (deficiency needs)

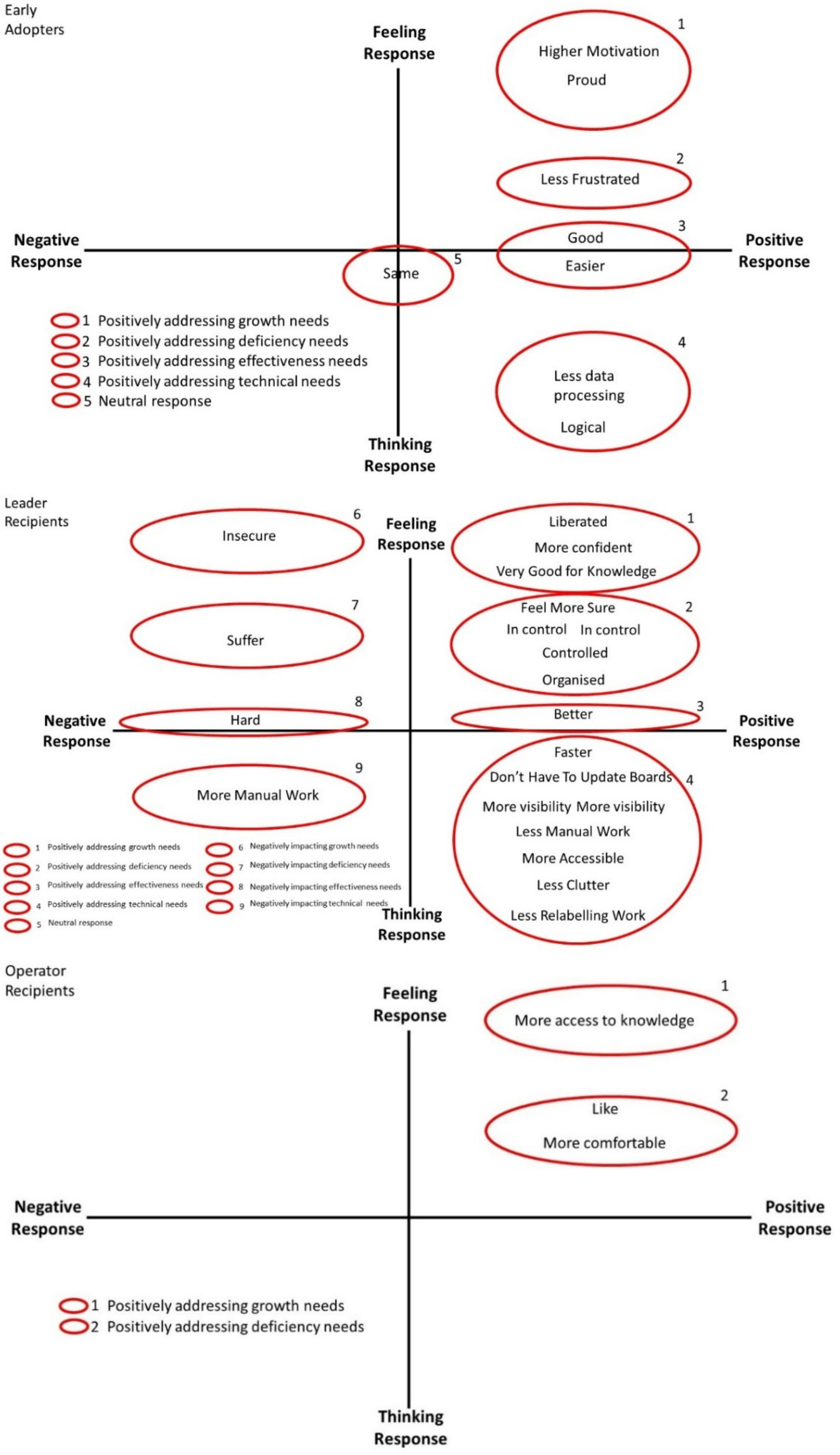

**Figure 5 Caption.** Reported feelings about the LI4.0 integration.
**Figure 5 Alt Text.** Responses are categorized as positive or negative, as well as more feeling-based (limbate response) or thinking-based (cortex response) and are, hence, displayed in four quadrants.

To delve further into the respondents' view of the performed work, they were asked about their perception of the integration transition. Two positive threads appear. First in technology terms, starting with the 'standardization of data' and 'real-time communications', which allow for 'automatic data capture'. 'Automatic data capture' is the key fulcrum of the loop because this allows for the 'stability of data', 'better measurement', 'the removal of manual readings', 'reduction of paperwork', 'the availability of documents' and 'visualization on one (digital) screen'. In turn, these benefits lead to 'fewer human errors', 'reduction of paper storage', 'better spotting of real issues', 'everything easier to see' and, generally, a 'better view of production'. These benefits permit a 'more organized workplace', 'quicker interventions', 'predictive warnings' and 'anticipation of problems'. The two main outcomes from the employee perspective are the 'improved time utilization of gaps' (estimated at about 1 hour a shift) and an 'improvement in people's work'. The second thread is clearly distinct and concerns people's learning and development. There are only three benefits identified in this area: 'all the learning I have achieved', 'people have developed by acquiring new skills' and 'now everyone can do it'.

There are far fewer negative points than positive ones. Hence, we asked each respondent about their worst aspect of their post digitization work to delve more deeply into the less beneficial areas and what could have been done better. Unlike the positive points, there is no obvious thread. The most cited worst point is connectivity, which is clearly vital to any I4.0 technology integration. In this case, there are internal slow and lost connectivity issues, both with the new solutions, and also between these and other existing solutions. These issues also occur with connectivity to external cloud-based solutions, which is partly due to 5G not being available. The second most cited issue concerns the usability of some solutions. This point would, of course, have been worse for this model site than for subsequent roll out sites because many of these teething issues are likely to have been corrected.

A range of other worse points were raised by two or three respondents, including configuration issues, issues caused by inventory data accuracy, the fact that many solutions were being simultaneously implemented, solution failures (once again, which may be less of an issue in further site roll outs), time-fixing problems, frustrations that implementation was not going faster, as well as two more human issues about the availability of human resources and training.

Further questioning of opportunities involved asking what could have been done better when there was a high degree of consensus. By far the most cited is that of training, in stark contrast to the minor learning and development thread discussed above. This suggests that this area could have been done better. Much of this training is undertaken digitally rather than in a lean coaching style. The second major opportunity is quicker problem solving, which is always likely to be the case in this first-of-its-type model plant application. This is probably highlighted because the people in this lean plant would expect to be able to quickly solve problems. However, as many felt they had neither sufficient training nor access to external resources, they were likely to feel frustrated and have a sense of lost autonomy. Several responses suggest that a more step-by-step approach for different solutions is necessary, as is making more human resources available. These points lie mostly in the human area.

The final respondent enquiry area includes views about people's general enthusiasm for lean digitization (Table 4). All three respondent groups obtain a very high positive response: 6.80 for early adopters, 6.00 for operator recipients and 5.47 for leader recipients. A high score might have been anticipated for early adopters seeing that they 'own' the program. Perhaps the high score for operator recipients is more surprising, with the slightly lower score for leader recipients explained by the hard work required by the office staff, supervisors and gaps during the implementation period.

It is notoriously difficult to measure the tangible results of any change and if they could have happened without change. However, estimates are provided in Table 6, which shows a reduction in inventory, inventory variance and scrap, and in labor savings, in the order of 0.68% - 2.5%. The largest labor saving is that provided by eLean because this saves gap leaders about 1 hour a shift, and this time is reused for improvement activity. These figures should be considered in the context of a very mature lean site where lean has already been employed before digitization and, hence, many of the easy savings from lean have already been made. In addition, intangible benefits are included: an improvement in maintaining standards, creating a modern digital factory that can help to retain and attract employees, and a modular setup that can allow new solutions to be added. However, there is a slower rate of improvement of ideas from operators because the digitized improvement system is not readily accessible to them. Hence they set up e-booths on the shop floor for operators. This does not, however, increase the significant number of improvement ideas because operators like to fill in old paper documents in their own time during breaks. As a result, the site has reinstated paper forms for operators that are then digitized by gap leaders.

**Table 6.** Estimates of quantitative savings at the studied site.

| | | **Stock** | | | **Non quality cost savings** | | | **Cost savings** | | |
|---|---|---|---|---|---|---|---|---|---|---|
| **Process** | **Solution** | **Raw material** | **Work in progress** | **Finished goods** | **Exception transport** | **Inventory deviation** | **Scrap** | **Direct labor** | **Logistics labor** | **Indirect labor** |
| **MPS** | **CMfg** | 5% | 0% | 1% | 0% | 0% | 0% | 0% | 1% | 1.7% |
| **Leveling** | **DPP** | 0% | 0% | 6% | 0% | 0% | 0% | 0% | 0.3% | 1.7% |
| **Pull Production and quality monitoring** | **eKanban** | 0% | 0% | 6% | 0% | 10% | 0% | 0.16% | 0% | 1.7% |
| | **MES/ BI/SPC** | 0% | 0% | 0% | 0% | 2% | 1% | 0.25% | 0% | 0.8% |
| **Shopfloor improvement** | **eLean** | 0% | 0% | 0% | 0% | 0% | 1% | 0.27% | 0.2% | 0.8% |
| | **Total (%)** | **5%** | **0%** | **7%** | **0%** | **12%** | **2%** | **0.68%** | **1.5%** | **2.5%** |

## 5. Discussion

The existing literature about LI4.0 remains limited and particularly lacks in-depth, holistic studies. This paper contributes by presenting a human-centered LI4.0 case study in the model plant of an automotive first-tier supplier. The findings of the formulated RQs are discussed below.

**RQ. What should be considered to integrate the triad of lean, I4.0 and human?**

The company under study already had an advanced lean approach at both the group level and in the plant under study. It, therefore, adopted LI4.0 by digitizing its lean approach, a route followed by most other organizations (Rosin et al., 2019). The work started at the group level before being adopted at three model sites, one of which is examined here. From this case, we cannot conclude that this is the only effective way, but we can confirm that, at least in this case, it is effective and the company’s preferred way. Regarding the human perspective, the case highlights several social mechanism but that can be categorized into three main interactions: (i) human-centered technology utilization that serves employee needs, reinforce their autonomy and organizes cooperation; (ii) short-term and mid-term collective behaviors protected by HR, managers and peer communications and (iii) competency development through training, coaching and mentioned collective behaviors.

**RQ1. What mechanisms and/or tools foster this integration?**

Based on the company's approach, the integration of human-centered LI4.0 is facilitated through several key mechanisms. The company XPS is the foundation, employing a house of excellence framework to promote a shared work culture and decentralized organizational principles. XPS highlights people development, stable working conditions, and sustainability, fostering engagement, competency development in autonomous teams, and environmental stewardship. Concurrently, the eLean suite provides digital tools for real-time production monitoring, problem-solving support, visual standard work routines, and guided improvement activities, enhancing operational efficiency and employee engagement. The digital manufacturing platform integrates I4.0 technologies such as cloud-based planning, production scheduling, and real-time sequencing systems (eKanban), optimizing production efficiency and quality management through automation and data-driven insights. Education and training initiatives complement these efforts, ensuring workforce readiness and alignment with continuous improvement goals across all organizational levels. Results shows that all dimensions of the Self-Determination Theory (Ryan & Deci, 2000) were taken into account while designing the infrastructure to support the adoption. Together, these mechanisms support a holistic approach to integrate operational, social and technological perspectives of human-centered LI4.0, enhancing organizational effectiveness and sustainability.

**RQ2. What are the main implementation results of this integration?**

The effectiveness of implementation can be seen by applying the proposed human-centered LI4.0 conceptual model (Figure 6). This breadth of results helps to address the gap highlighted by Pagliosa et al. (2021). We present the outcomes within the three perspectives that we use in this paper, although there is evidently a marked overlap and interaction among these perspectives. However, it is only by looking through all three that can we gain this triangulated viewpoint. In many ways, the addition of the social perspective (Romero et al., 2020) is the most insightful, as it highlights that site employees at all levels perceive significant improvements in lean practices, their work experience, and their overall attitude toward digitization.

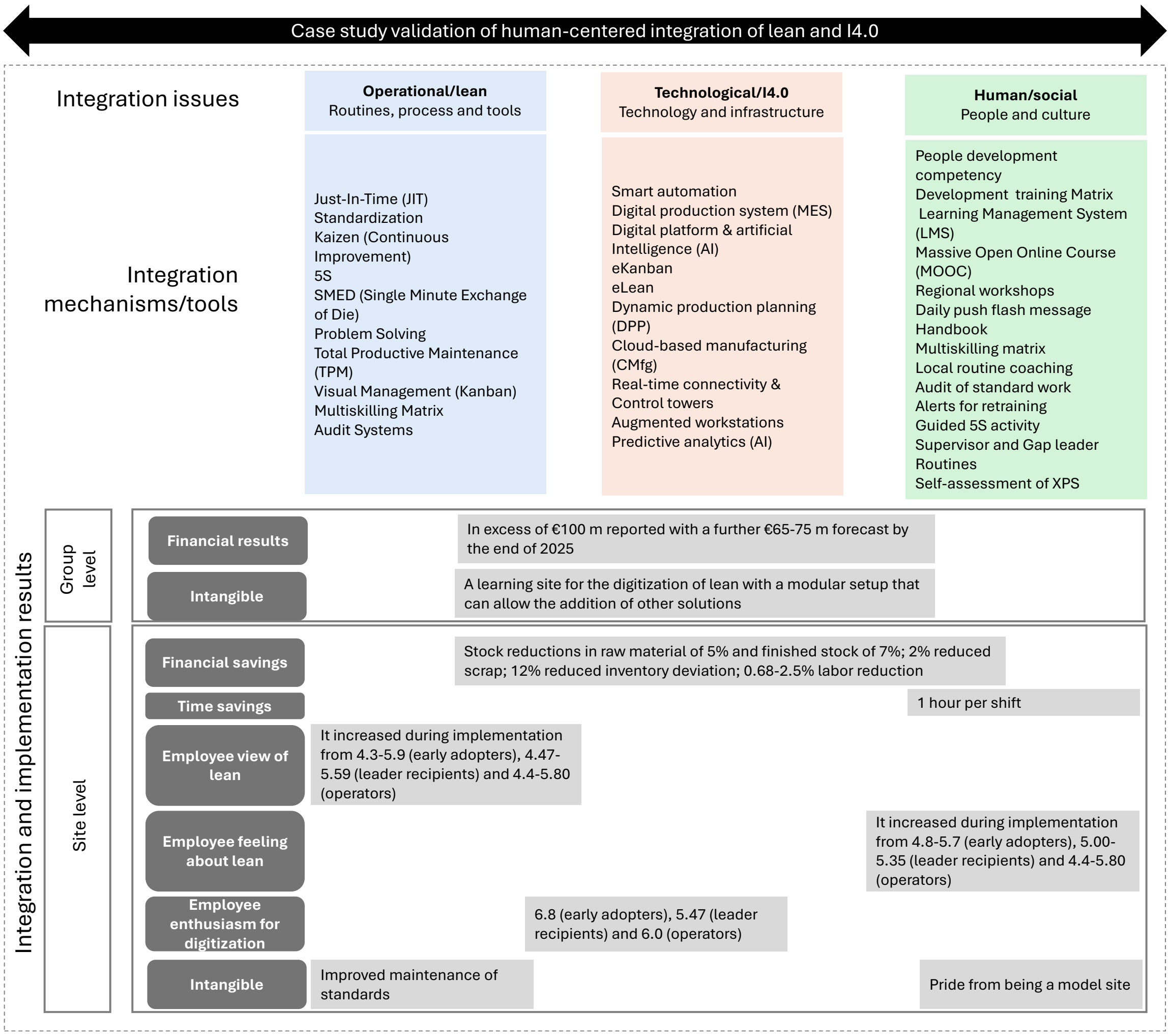

**Figure 6 Caption**. Human-centered LI4.0 case study model.

**Figure Alt Text**. Human-centered LI4.0 case study model shows benefits at both the group and site levels.

The human-centered LI4.0 conceptual model of the case study provides a wealth of evidence of what is happening and why. Based on this, we summarize the results in a CLD (Figure 7). This approach allows us to identify the relations among variables, which are not visible at first glance or are mediated by other factors, to appear visually evident (Cimini et al., 2023). What we have seen is advanced lean implementation that puts technology to very good use and closely to replicate the pre-existing lean state.

The overall triple perspectives approach uncovers a complex web of interactions, which are mainly positive for explaining the positive implementation result. At the group level, this involves 10 positive nodes (by creating 23 positive links with other causal nodes) and three negative nodes (by creating four negative links). This shows that the impact of work at the group level is overwhelmingly positive. At the site level, there are 16 positive nodes (by creating 22 positive links) and seven negative nodes (by creating six negative links). This shows that the impact of work at the site level is also overwhelmingly positive. Further positive results can be anticipated in the future as work in the AI area adds machine intelligence to the already employed human one. The negative nodes at both the group and site levels are also anticipated to be addressed by the company.

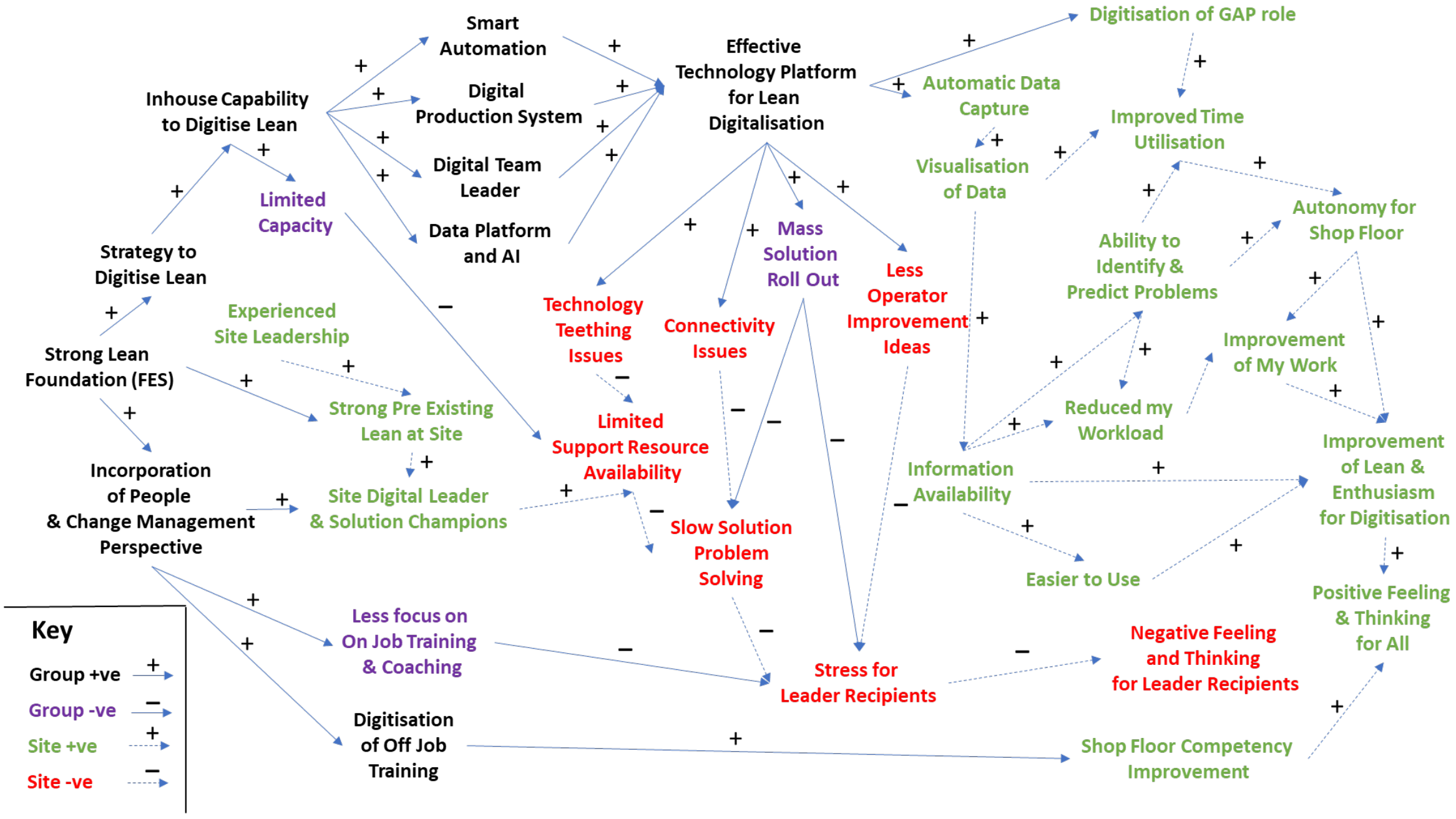

**Figure 7 Caption.** The closed-loop diagram (CLD) for model plant implementation. **Figure 7 Alt Text**. A CLD to identify the relations among the main variables of the model plant implementation.

The 26 positive nodes prove to be strong enablers in the case with the 10 negative nodes, and also prove to be inhibitors. At the group level, success is linked with a very solid lean foundation, which is enabled by a wide range of technological factors. In terms of opportunities at the group level, two lie in the people area and one is technological. The two people-related areas are about increasing capacity to help not only to support the site (and the subsequent roll out), but also site staff training and coaching. Both are about the development competency. The first relates more to the support of early adopters and leader recipients, and the latter is more related to leader recipients. These are, therefore, important competencies that add to the requirement in this area (Ansari et al., 2018).

Operators' 'virtual classroom' training seems to be most effective. This position can be explained by the reference to the established 10:20:70 learning theory (Hines & Butterworth, 2019) that, for optimum results, suggests that 10% learning should be classroom-based, 20% should be on-the-job, and 70% with subsequent manager and/or coach follow-up. The 10% element is particularly important for developing explicit knowledge (as required by operators) at the tool level, with the 20% and 70% elements being especially relevant for developing tacit knowledge (as required by early adopters and leader recipients) (Bortolotti et al., 2015). We can conclude that this was forthcoming during the time when on-site group support was available. However, it was lacking after this time, which brought about a range of negative impacts, especially on leader recipients. We can also conclude that the ongoing 'virtual classroom' was insufficient and required the addition of more physical coaching. All this forms the first of our proposed limits to digitization.

The site also started with a very solid lean position, and most positives are due to the employed technology and its ability to visualize data in a form that could be readily used. This led to a wide range of positive people issues with both limbate and cerebral benefits because work not only helped shop floor workers to reduce their workload, but also led to improvements in autonomy and competence and, hence, improved their motivation (Ryan & Deci, 2000). This, therefore, enabled many of the surveyed employees to satisfy both growth and deficiency needs. The effect on relatedness was more neutral.

Most of the negative areas at the site level were people related to two areas in the technology area, namely 'technology teething issues', mainly caused by this being a pilot plant, and 'connectivity issues', as suggested by Yilmaz et al. (2022). There were clearly some areas of frustration for leader recipients in particular. These were concerned with speed, resource availability and the fact that operators were less engaged in raising problems ideas. All this caused stress and had a largely emotional effect on their well-being. The fact that even the advent of shop floor booths for operators suggestions did not work well meant that this use of digital technology falls into our second limit to digitization because it cut across operators' autonomy (Rosin et al., 2019) and work routines (Pagliosa et al., 2021). It consequently resulted in them producing fewer ideas.

## 6. Conclusions

In this paper, we take a three-perspectives approach to propose an integrated human-centered LI4.0 conceptual model and, based on this, view advanced and highly successful LI4.0 model plant implementation. This yields wide-ranging findings that are summarized in a CLD. We believe that the integrated human-centered LI4.0 conceptual model and its implementation in this holistic case adds to the scare literature about the subject. It also provides valuable insights into several key gaps in the extant literature, such as the addition of a social perspective, identification of enablers and barriers, implementation results and the CLD. This work particularly addresses critical gaps, such as the need to better understand the implementation sequence of LI4.0, and the alignment between lean practices and I4.0 technologies through a human-centered approach. Additionally, the social perspective highlights important areas of interaction between humans and technology, and emphasizes the relevance of social engagement throughout digital transformation processes. These aspects are frequently overlooked in most LI4.0 literature works and in the nascent I5.0 literature, which underlines the importance of integrating the social dimension into a holistic understanding of LI5.0.

Regarding theoretical implications, we have shown the importance of the social perspective and we, thus, propose that future research should adopt a triple-perspective approach, as demonstrated in this work. This will increase researchers' focus and draw researchers from a wider disciplinary background. Through our CLD we have demonstrated the complexity of interactions and how a successful human-centered LI4.0 model can be implemented.

Managerial implications are oriented to the fact that the integrated human-centered LI4.0 model, combined with the case study approach, has proven to be very successful and serves as a valuable role model. The company achieves a high level of lean thinking at the group level and in the model plant before starting the digitization process, which has been shown to be invaluable. We also recommend placing more emphasis on understanding how affected employees feel and think before, during and after implementation, with particular attention paid to softer aspects like training, coaching and change management.

This work has several limitations. We rely on the literature review, and the data and viewpoints provided by individuals from the company under study, which we sought to triangulate and validate. However, full triangulation was not completely possible, especially with the financial data that we received. In addition to this, we would caution that this is only a single case study and the generalization of what we found should be carefully addressed. The journey we have seen here is unique to the company, and indeed to this particular site. Nevertheless, we believe that many of our observations and propositions may have broader applicability.

We encourage future research to build upon our human-centered LI4.0 conceptual model by conducting further cases-based studies across different sectors, countries and organization sizes. It would be particularly useful to study longitudinal cases to, for instance, analyze what happens at our case site in forthcoming years, particularly in terms of sustainability and the effect of AI technology. For our case company, it would also be useful to further explore the group program in terms of its multi-plant roll out.

**Acknowledgments**

The research that has led to the present results has received funding from the MCIN/AEI/10.13039/501100011033 and by European Union Next Generation EU/PRTR with grant agreement PDC2022-133957-I00 (CADS4.0-II); and the Regional Department of Innovation, Universities, Science and Digital Society of the Generalitat Valenciana entitled "Industrial Production and Logistics Optimization in Industry 4.0" (i4OPT) (Ref. PROMETEO/2021/065) and the European project HORIZON-CL4-2024-TWIN-TRANSITION-01-03 with grant number 101177842 (UniMaaS).

**Data availability statement**

The authors confirm that the data supporting the findings of this study are available in the article and in Appendices included as a supplementary file.

**Disclosure statement**

No potential conflict of interest is reported by the authors.

## Appendix A: Semi-Structured Interview Questions

1. Please can you tell me your name and role at your company?
2. How many people are there in your team?
3. How long have you worked at the site?
4. How have you been involved in the lean work at the site?
5. What has been your role in the digitization (Lean Industry 4.0) of lean at the site?
6. Which solutions do you use?
7. What is your view of the effectiveness of Lean before digitization (1 = very bad – 7 = very good)?
8. What is your view of the effectiveness of Lean Industry 4.0 work (1 = very bad - 7 = very good)?
9. Can you explain this answer?
10. What have been the results in your area?
11. How did you feel about your work before digitization? (1 = very bad - 7 = very good)?
12. How did you feel about your work after digitization? (1 = very bad - 7 = very good)?
13. Can you explain this answer?
14. What have been the best aspects of work?
15. What have been the worst aspects of work?
16. What could have been done better?
17. What is your view of the new technology?
18. What level of enthusiasm do you feel for Lean Industry 4.0 work (1 = very bad – 7 = very good)?
19. Please describe one critical event that increased your enthusiasm for Lean Industry 4.0 work
20. Please describe one critical event that decreased your enthusiasm for Lean Industry 4.0 work

**Appendix B: Digital Lean Plant Survey**

1. What is your role at the site?
    a. Senior manager
    b. Manager
    c. Key user
    d. Supervisor
    e. GAP leader
    f. Other indirect labor
    g. Operator
2. What has been your role in the digitization of lean at the site?
3. Which of the following digital solutions do you use? [multiple answers possible]
    a. MIP
    b. PREACTOR
    c. NEO
    d. DMC
    e. MLEAN
    f. None of these
4. What is your view of the effectiveness of Lean before digitization (1 = very bad - 7 = very good)?
5. What is your view of the effectiveness of Lean digitization work (1 = very bad - 7 = very good)?
6. Can you explain any change in your answer to the last two questions?
7. What have been the results in your area?
8. How did you feel about your work before digitization? (1 = very bad - 7 = very good)?
9. How did you feel about your work after digitization (1 = very bad - 7 = very good)?
10. Can you explain any change in your answer to the last two questions?
11. What has been the best aspect of work?
12. What has been the worst aspect of work?
13. What could have been done better?
14. What level of enthusiasm do you feel for Lean Digitization work (1 = very bad - 7 = very good)?